\documentclass[useAMS,usenatbib]{mn2e}
\usepackage{graphicx}
\usepackage{epstopdf}
\usepackage{epsfig}
\usepackage{dcolumn}
\usepackage{bm}
\usepackage{amssymb}
\usepackage{cite}
\usepackage{multirow}
\usepackage{color}

\newcommand{\hompc}{\,h\,{\rm Mpc}^{-1}}
\newcommand{\mpc}{\,{\rm Mpc}}

\newcommand{\mpcoh}{\,h^{-1}\,{\rm Mpc}}

\def\be{\begin{equation}}
\def\ee{\end{equation}}
\def\ba{\begin{eqnarray}}
\def\ea{\end{eqnarray}}

\begin{document}

\title{Cosmic distances probed using the BAO ring}

\author[Cristiano G. Sabiu, Yong-Seon Song]{Cristiano G. Sabiu, Yong-Seon Song\\
Korea Astronomy and Space Science Institute, Daejeon, 34055, Korea\\}

\date{} 

\maketitle

\label{firstpage}

\begin{abstract}
The cosmic distance can be precisely determined using a `standard ruler' imprinted by primordial baryon acoustic oscillation (hereafter BAO) in the early Universe. The BAO at the targeted epoch is observed by analysing galaxy clustering in redshift space (hereafter RSD) for which a theoretical formulation is not yet fully understood, and thus makes this methodology unsatisfactory. The BAO analysis following a full RSD modelling is contaminated by systematic uncertainties due to a non--linear smearing effects such as non--linear corrections and by random viral velocity of galaxies. However, the BAO can be probed independently of RSD contamination using the BAO peak positions located in the 2D anisotropic correlation function. A new methodology is presented to measure peak positions, to test whether it is also contaminated by the same systematics in RSD, and to provide the radial and transverse cosmic distances determined by the 2D BAO peak positions. We find that in our model independent anisotropic clustering analysis we can obtain about $2\%$ and $5\%$ constraints on $D_A$ and $H^{-1}$ respectively with current BOSS data, which is competitive with other analysis.

\end{abstract}

\section{Introduction}
Precise measurement of cosmic distance has become a key targeted observable for probing the cause of cosmic acceleration. The cosmic acceleration has been confirmed by many observations \citep{Riess1998,Perl1999}, and our understanding of the universe has undergone a revolution with this new discovery. The standard model of the universe needs to be altered to either contain the yet unknown dark sector energy components, or to modify gravitational physics at large scale. The cosmic distance probes will provide us with the evolution of cosmic expansion from the epochs in which the cosmic expansion starts to be accelerated, and with the clues to include dark materials or dark gravity in order to complete the standard model of the universe.

Cosmic distances are observed by cosmic parallax, standard candles and standard rulers, but most methods suffer from their own systematics. Among all of these diverse approaches, the standard ruler method is considered the most risk--free technique to probe cosmic distance. The acoustic peak structure of the baryon--photon fluid in the early universe is developed by the tension between gravitational infall and radiative pressure and imprinted on the last-scattering surface. The fossil record remain imprinted on the large-scale galaxy clustering even today \citep[hereafter BAO ][]{Blake:2003rh,Seo:2003pu,Eisenstein:2005su}. The characteristic scale of the BAO can be used as a standard ruler, which enables us to determine geometric distances for high-$z$ galaxies to great precision. Previously geometric distances were determined by analysing simultaneously the clustering anisotropies over the BAO scales. The galaxy clustering at the targeted area is observed in the redshift space (hereafter RSD), which is contaminated by non--linear physics. The transformation between real and redshift space is intrinsically non--linear, in that the density perturbation and velocity fields are coupled together and evolve nonlinearly. The factorized formulation has been proposed to achieve the RSD theoretical model, which turns out to be the combination of non--separable linear squeezing effect and non--linear smearing effects caused by those higher order polynomials. The small scale distortion along the line of sight is caused by the randomness of the velocity field, called the Finger of God (hereafter FoG) effect. This effect is non--perturbative, and there is little clue of its exact theoretical form neither in scale nor in time. All of these systematics are combined in the full RSD analysis, which make lowers our confidence on the cosmological constraints from the RSD effect beyond a conservative limited scale at the quite linear regime. 

However, the characteristic BAO distances can be probed alternatively, being independent of the detailed knowledge of galaxy clustering in redshift space, by making measurements of the oscillation scale while ignoring or otherwise marginalising the overall shape.
In recent efforts several works have targeted the BAO scale using the 1D isotropic density power spectrum and correlation functions \citep{2012MNRAS.427.3435A, 2014MNRAS.441...24A} while ignoring or otherwise marginalising out the overall shape of the clustering signal. The BAO scale has been shown to be quite insensitive to non-linear physics and other systematics \citep{2014PhRvD..89j3541S}, however for near future experiments e.g. DESI some of these systematics will have to be understood or controlled to a greater degree \citep{2014arXiv1410.4684P}.

The anisotropies of the clustering pattern of the galaxy distribution can also arises from any discrepancy between the true underlying cosmological model and the model used to convert the redshift and angular position of each galaxy to the co-moving radial and transverse distances. This is the so-called Alcock-Paczynski (A-P) effect \citep{AP1979} and with a prior knowledge of the characteristic scale of the BAO, the Hubble parameter $H(z)$ and angular diameter distance $D_A(z)$ at the high-$z$ galaxies can be separately measured \citep{2003PhRvD..68f3004H, 2014PhRvD..89j3541S}.




There have been several methods proposed that apply the AP test to the large scale structure by measuring the clustering of galaxies \citep{Ballinger1996,Matsubara1996,2015MNRAS.450..807L}, symmetry properties of galaxy pairs \citep{Marinoni2010}, and cosmic voids \citep{Ryden1995,LavausWandelt1995}. Among them, the method of galaxy clustering has been widely used to constrain cosmological parameters \citep{Outram2004,Padmanabhan:2008ag,2009MNRAS.399.1663G,Blake:2011en,Blake2011,ChuangWang2012,2014arXiv1407.2257S, Linder2013, Beutler2013, Reid2012,2013MNRAS.429.1514S,Jeong2014,2014MNRAS.440.2692S,2015MNRAS.451.1331R}. The main caveat of this method is that, because the radial distances of galaxies are inferred from redshifts, AP tests are inevitably limited by redshift-space-distortions (RSD), which leads to apparent anisotropy even if the adopted cosmology is correct \citep{Ballinger1996}.


In this work we will introduce our anisotropic (2-dimensional) galaxy clustering statistic and show how in an ideal case, free of systematics, this can be used to locate the BAO peak location and thus obtain information on $D_A$ and $H$ independently. We also show that this methodology is robust against the systematics of unknown bias and undetermined shape of the underlying power spectrum. However, as we will show later, the unknown  FoG effect causes an anisotropic shift in the BAO location which induces a degeneracy been it and Hubble parameter $H(z)$. Thus we proceed to model this shift and marginalise out its effects reducing the systematic bias.  This methodology is robustly tested using Monte Carlo simulations of the galaxy distribution where we show that we can reliably recover the fiducial input cosmological parameters.

The outline of this paper proceeds as follows.  In \S 2 we briefly review the Alcock-Paczynski effect in a cosmological context and describe the statistical methods we will employ. 
In \S 3 we investigate the BAO peak structure and its sensitivity to various systematic uncertainties.
In \S 4 we test our methodology on simulated mock galaxy catalogs and derive cosmological parameter constraints.
We conclude in \S 5.

\section{determination of geometrical distances using the BAO ring}



The observed galaxy clustering anisotropies are plagued by systematic uncertainties which causes the measured geometric distances through BAO feature unreliable compared to other distance probes of the Universe. When the galaxy clustering is viewed from the redshift space, the cosmological density and velocity fields couple together and evolve nonlinearly. In addition, the mapping formula between the real and redshift space is intrinsically nonlinear. These nonlinearities prevent us from inferring the linear coherent motion from the redshift space clustering straightforwardly. Although an accurate theoretical model for the redshift distortion was proposed in, there is no confirming statement whether the measured distances are immune from the complexity arising from all different types of theoretical models to explain the cosmic acceleration.

We propose that the geometrical distances can be determined with exploiting BAO feature as minimal as possible. In the context of standard cosmology, the BAO peak structure is known by analysing cosmic microwave background (hereafter CMB) anisotropy produced before the last scattering surface. The primordial BAO feature is  probed in precision by CMB experiments. Then we claim that there is a unique correspondence between the measured primordial BAO feature by CMB and the observed geometrical BAO signature by galaxy clustering survey, which is nearly independent of other cosmological uncertainties. 

\begin{figure}
\centering
\includegraphics[width=1.0\columnwidth]{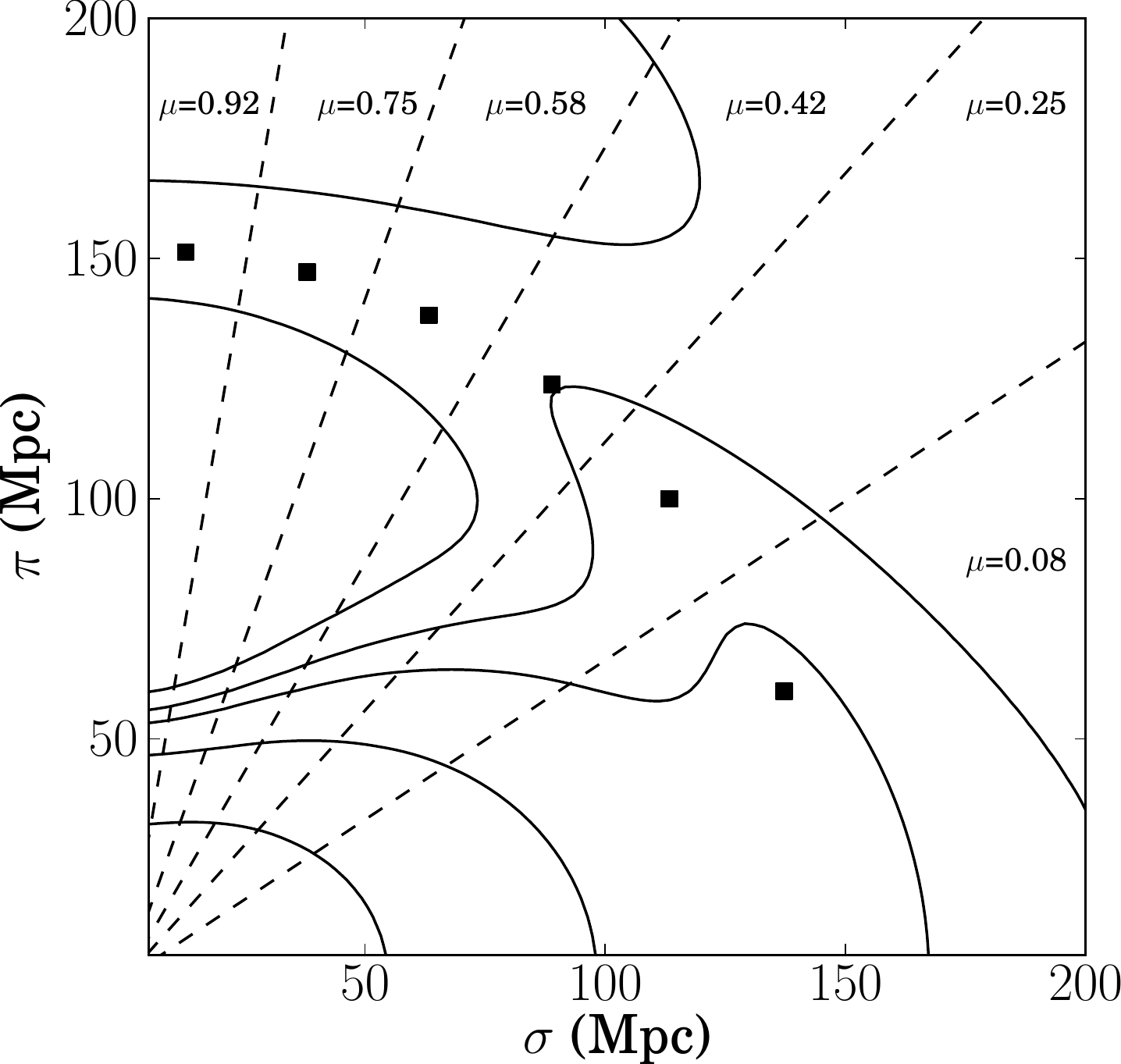}
\caption{\label{fig:sigpi_contour} The correlation function of $\xi(\sigma,\pi)$ is presented with equi-clustering contours at  $\xi(\sigma,\pi)=  [0.1, 0.02, 0.005, 0.001, -0.003]$. Dash lines represent boundaries of different wedges which are divided by linear $\mu$ scale from $\mu=0$ to 1. The squared points denote the BAO peak of one-dimensional correlation function of $\bar\xi_\mu(s)$ at each wedge, which are located on the great circle BAO ring connecting tip points of all contours of $\xi(\sigma,\pi)$.}
\end{figure}

The geometrical signature of primeval BAO is precisely observed through galaxy clustering correlation function $\xi(\sigma,\pi)$, where $\sigma$ and $\pi$ denote the transverse and radial distances. The two point correlation distance $s$ is defined by $s\equiv\sqrt{\sigma^2+\pi^2}$. The clustering patterns are different between inner and outer contours. The monotonous anisotropic shape is observed with inner contours at $s\la 100\mpc$. The squeezed clustering pattern along the radial direction is determined by the competition between monopole and quadrupole amplifications which are caused by density and velocity fluctuations respectively. BAO peak structure is observed at outer contours encompassing to cross $s\sim 150 {\rm Mpc}$. The primordial wiggles on the spectra turns out continuous peaks along the BAO ring at $s\sim 150\mpc$ which takes an important role to measure distances. The appearance of BAO ring is presented in Fig.~\ref{fig:sigpi_contour}. 

In our previous work, the full anisotropy analysis is exploited to measure distances. The BAO ring is observed to be invariant by the variation of structure formation when the broadband shape of spectra is previously determined by CMB experiments. Thus we are able to probe distances after marginalising over all structure formation parameters, such as the density \& velocity growth functions, the galaxy bias and the random velocity effect. Although it is a thorough methodology, the measured distances can be biased by unknown clustering systematics. 

Inspired by the invariance of the BAO ring, we suggest a new methodology to determine it, independent of any knowledge of the galaxy clustering. In practice, the observed spectra are estimated using referenced coordinates, not true coordinates, because the transformation rule between the observed (Ra, Dec, z) and the cartesian coordinates of distance measures is unknown. The fiducial cosmology is used to convert the observed (Ra, Dec, z) into comoving Cartesian coordinates. The degree of spatial distortion from this fiducial coordinate can be described by the following quantities,
\begin{equation}\label{eq:stretch}
 \frac{[\Delta \pi/\Delta \sigma]_{\rm fid}}{[\Delta \pi/\Delta \sigma]_{\rm true}} =
  \frac{[D_A(z)H(z)]_{\rm true}}{[D_A(z)H(z)]_{\rm fid}},
\end{equation}
where $\Delta \pi$, $\Delta \sigma$ are the angular and radial sizes of the objects,
and ``true'' and ``fid'' denote the values of quantities in the true cosmology and fiducial cosmology.
$D_A$ and $H$ are the angular diameter distance and Hubble parameter, respectively. 
These geometrical quantities can provide us with information on the particular energy components of the Universe and in the specific case of a model with constant dark energy equation of state.
However, if we independently determine the BAO ring, which spans both transverse and radial cosmological coordinates, it can be exploited to probe the distance measures of $D_A$ and $H^{-1}$.

We proceed to construct an estimator to locate the BAO peaks of $\xi(\sigma,\pi)$ represented by the squared points in Fig.~\ref{fig:sigpi_contour}. The observed distance between two galaxies $s$ is defined assuming a fiducial or reference cosmological model, 
and the observed cosine of the angle the pair makes with respect to the line of sight, $\mu$ is given by
\begin{equation}
\mu=\frac{\pi}{s}\,.
\end{equation}
The estimate of these separations is dependent on the assumed cosmology model. 
We estimate the 2-point correlation function (2PCF) in redshift-space and in the anisotropic $s,\mu$-decomposition transformed from $\sigma$, $\pi$ coordinates. 
The correlation functions are calculated using the ``Landy-Szalay" estimator, 
\begin{equation}\label{eq:LSesti}
\xi(s,\mu)=\frac{DD(s,\mu)-2DR(s,\mu)+RR(s,\mu)}{RR(s,\mu)},
\end{equation}
where $DD$ is the number of galaxy--galaxy pairs, $DR$ the number of galaxy-random pairs, and $RR$ is the number of random--random pairs, all  separated by a distance $s\pm\Delta s$ and angle $\mu\pm\Delta\mu$. The pair counts are normalised since we use 20 times as many randoms and data point to reduce shot noise contributions to the correlation estimation. 

The $\mu$ is divided into 6 bins with $\Delta \mu=1/6$ as shown in Fig.~\ref{fig:sigpi_contour}. The correlation function, $\xi(s,\bar\mu)$, where $\bar\mu$ denotes the median $\mu$ at each $\mu$ bin, is estimated by the fitting function, $\bar\xi_{\mu}(s)$, which is given by,
\begin{equation}
\bar\xi_{\mu}(s)\times s^2=A.s^2 + B.s+ Ee^{-(s-D)^2/C} + F, \label{eq:model}
\end{equation}
which is just a quadratic function to mimic the overall shape of the correlation function and a gaussian to model the BAO peak.
In our work the focus will be on constraining the scale parameter, $D$, as a function of the anisotropy angle, $\mu$. The measured $D(\mu)$ represents the radial distance of the peak location at the given $\mu$ bin which is presented at the squared points in Fig.~\ref{fig:sigpi_contour}. In this process the other 5 parameters will be marginalised over.

\begin{figure}
\centering
\includegraphics[width=1.0\columnwidth]{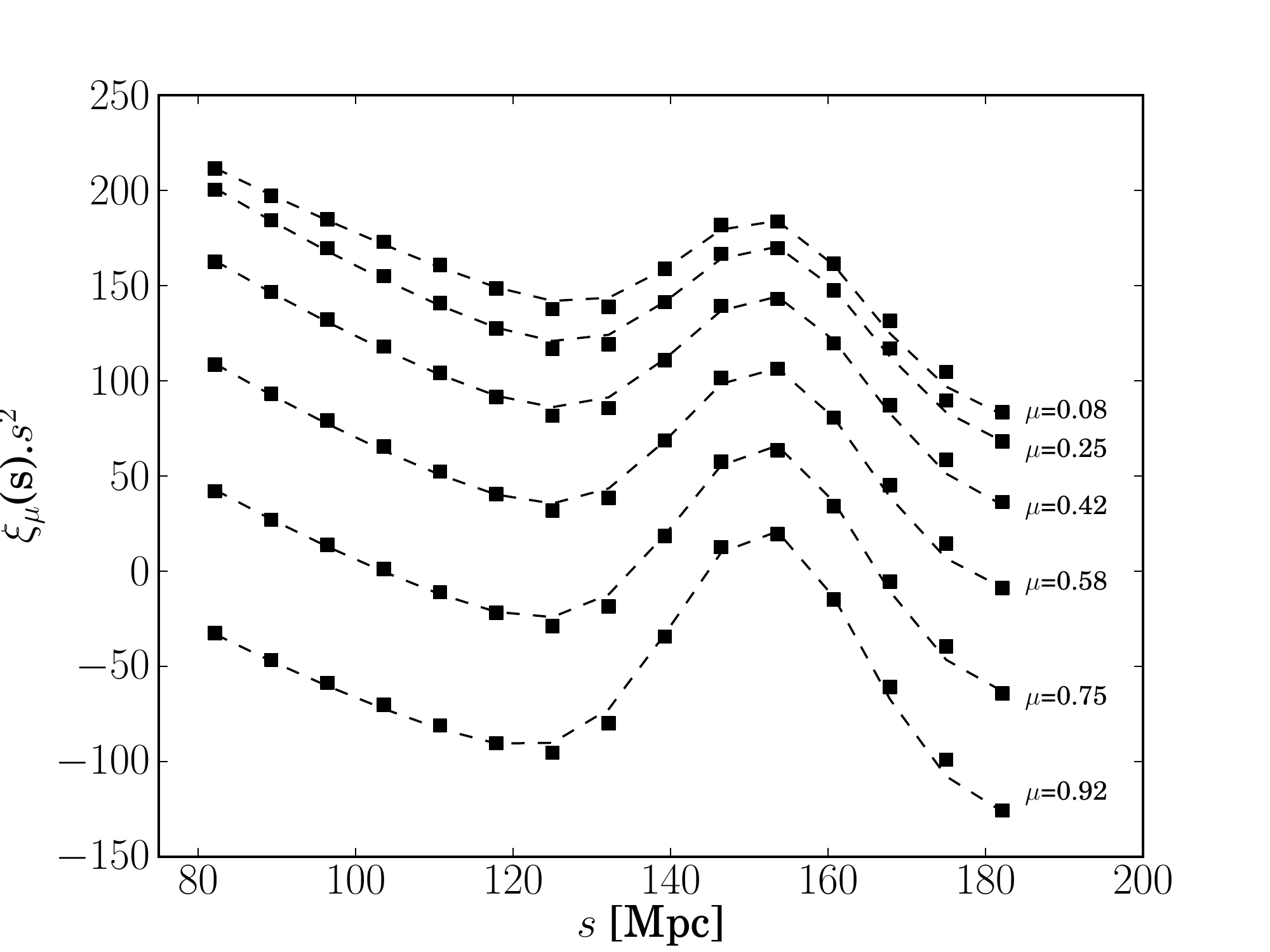}
\caption{\label{fig:xi_curves} The projected one-dimensional correlation function $\bar\xi_\mu(s)$ at each wedge is presented as black solid points. The correspondent $\mu$ for each curve is given by $\mu$=0.92, 0.75, 0.58, 0.42, 0.25, 0.08 from the top to bottom. The  dash curves represent the best fit curves defined by Eq.~\ref{eq:model}}
\end{figure}

We simulate the observed $\xi(s,\mu)$ using a theoretical linear correlation function. We then proceed to fit the individual $\bar\xi_{\mu}(s)$ to this simulation
using the model in Eq.~\ref{eq:model} while considering $1\%$ measurement error and no covariance between points. However in \S\ref{sec:mocks} where we consider mock galaxy data we will consider realistic errors and the full covariance between the data. In Fig.\ref{fig:xi_curves} we show the 2PCF, $\xi(s)$ for various $\mu$ values. In all $\mu$-directions the BAO feature is clearly seen. The fitting was done using a 20,000 chain mcmc over the the range in scales $80\mpc<s<180\mpc$, sampled in 15 linearly spaced bins. The errors on the measurements were assumed to be small, ~1\% and uncorrelated.  

The BAO characteristic distances located at peak positions of $\xi(s,\mu)$ appear as a great circle in the $s$ and $\mu$ plane which can be described by BAO peak ellipse as,
\begin{equation}
D(\mu)=\frac{D_{\perp}D_{||}}{\sqrt{(D_{\perp}\mu)^2 + D_{||}^2 (1-\mu^2)}}\,,
\label{eq:elipse}
\end{equation}
here $D_{\perp}$ and $D_{||}$ are BAO characteristic transverse and radial distances respectively. The observed angular positions and redshift are transformed to the co--moving distance for measuring $\xi(s,\mu)$ of the underlying theoretical model. The anisotropy pattern is induced to the observed $\xi(s,\mu)$ by the apparent possible discrepancy between the fiducial and true cosmologies. The true BAO ellipse is recovered by an appropriate transformation between two different cosmologies, and both $D_{\perp}$ and $D_{||}$ are determined. Those measurements are correlated with cosmic distance determinations as,
\begin{eqnarray}
H^{-1}_{obs}&=&H^{-1}_{fid}\frac{D_{||,fid}}{D_{||,obs}}, \nonumber \\
D_{A, obs}&=&D_{A, fid}\frac{D_{\perp,fid}}{D_{\perp,obs}}.
\end{eqnarray}
However, for the above relations to be considered unbiased estimators, the extracted distances from the fiducial model must be free from systematic uncertainties associated with our incomplete knowledge of galaxy clustering. The systematics will be investigated in the next section.




\section{Test on contamination by systematics in determining 2D BAO}

The initial matter-energy fluctuations cause the source for baryon acoustic waves that propagated in the photon-electron-baryon plasma of the early universe. These acoustic waves travel, before they stop at the epoch of recombination. The initial density peak of excess of matter observed at last scattering surface remains at the galaxy clustering at late epoch. This distance provides a standard ruler to determine both transverse and radial distances. In practice, there are nuisance systematic uncertainties, which prevent us from directly accessing the primeval BAO signature. Those systematics are classified into uncertainties due to the biased galaxy clustering to dark matter fluctuations, the non--perturbative effect from random velocities, and the unknown theoretical model of dark energy at late time. The methodology has been extensively developed to probe primeval BAO structure after marginalising all those nuisance parameters. The measured distances through simultaneous determination of all others are proved to be useful for diverse classes of theoretical models, but the determination is still model--dependent. We are interested in the opposite direction in which we remove those systematics from our analysis, instead of adding it to the analysis.

\subsection{Fiducial model}
The fiducial distances of $D_A$ and $H^{-1}$ are provided by theoretical RSD modelling using the perturbation theory. In computing the RSD power spectrum, we need to properly take into account the effect of nonlinear gravitational evolution for the auto- and cross-power spectra $P_{XY}(k)$. Since the standard perturbation theory is known to produce ill-behaved expansion leading to the bad UV behavior, a consistent calculation of the correlation function should be made with an improved perturbation theory that includes appropriate UV regularization. Here, we apply the resummed perturbation theory called {\tt RegPT}~\citep{Taruya:2012ut}, and following the prescription described in~\citep{Taruya:2013my}, we compute the power spectra $P_{XY}(k)$, including nonlinear corrections up to the two-loop. 

As we can see from the top left panel of Fig.~\ref{fig:xi_sys}, there is a noticeable difference in the clustering patterns of $\xi(\sigma,\pi)$ between the linear and non-linear models. More quantitatively this leads to a shift in the distances $D_{||}$ and $D_{\perp}$ about a percentage level which are summarised in the tables below. 
\begin{center}
\begin{tabular}{l|lll}
Template & $D_{||}$ & $D_{\perp}$ &   \\
\hline
linear &  152.12 & 153.33 &   \\
RegPT & 153.33  & 154.54 &  \\
\hline
\end{tabular}
\end{center}
This shift nessecarily requires us to model the clustering using non-linear templates even at large scales near the BAO peak position. However, the BAO ring is observed at relatively large scales of about $150\mpc$ in which the perturbative theory predicts the dominant higher order non--linear corrections in precision. The BAO remains as a standard ruler as far as the non--linear corrections are predictable in this regime. Therefore the non--linear effect on the BAO peaks is not counted as a systematic in this work, and the theoretical $\xi(\sigma,\pi)$ is derived from RegPT theory.

The cosmic distances of $D_A$ and $H^{-1}$ are determined by the measured shifts of the BAO distances, $D_\perp$ and $D_\parallel$, from their fiducial values. The variation of measured $D_A$ and $H^{-1}$ is presented in the following subsections with respect to changes of all nuisance systematics. The $D_\perp$ and $D_\parallel$ defined in Eq.~\ref{eq:elipse} are measured to be $154.5\mpc$ and $153.3\mpc$ for the fiducial cosmology. Distances of $D_A$ and $H^{-1}$ are calculated from the fractional difference in measured $D_\perp$ and $D_\parallel$ for each systematic uncertainty case. The fiducial $D_A$ and $H^{-1}$ at $z=0.57$ are $1395.2\mpc$ and $3234.8\mpc$.

\subsection{Galaxy bias}

Although the theoretical model of galaxy clustering to dark matter distribution is unknown, galaxy bias can be described using the following phenomenological function which is given by,
\ba
b(k)=b_0\frac{1+A_2k^2}{1+A_1k}
\ea
where $b_0$ denotes the coherent galaxy bias, and $A_n$ denote the scale dependent bias parameters. The shape of $\xi(s,\mu)$ changes with the variation of $A_n$, but there is little effect on outer contours at $s>100 {\rm Mpc}$. The determination of $D$ indicating the location of BAO peaks is immune from the scale dependent bias systematics. Therefore we mainly focus on the effect by coherent bias variation in this subsection. The observed location of BAO peaks varies with different $b_0$, but we observe that the 2D BAO ring remains invariant. In the table below we show the effect of changing the linear  bias factor on the derived distance measures. Since the bias only alters the amplitude of the clustering signal we should not expect a shift in the BAO peak position. However we investigate the change in bias in the case that our minimal model can still fit the peak position without introducing any systematic variation due to inaccurate fitting. We find that values of $b_0=1.5$, 2.0, 2.5 all give consistent values of $D_{\perp}$ and $D_{\parallel}$ and induce subpercent level shifts in  $D_A$ and  $H^{-1}$.

\begin{figure*}
\centering
\includegraphics[width=1\columnwidth]{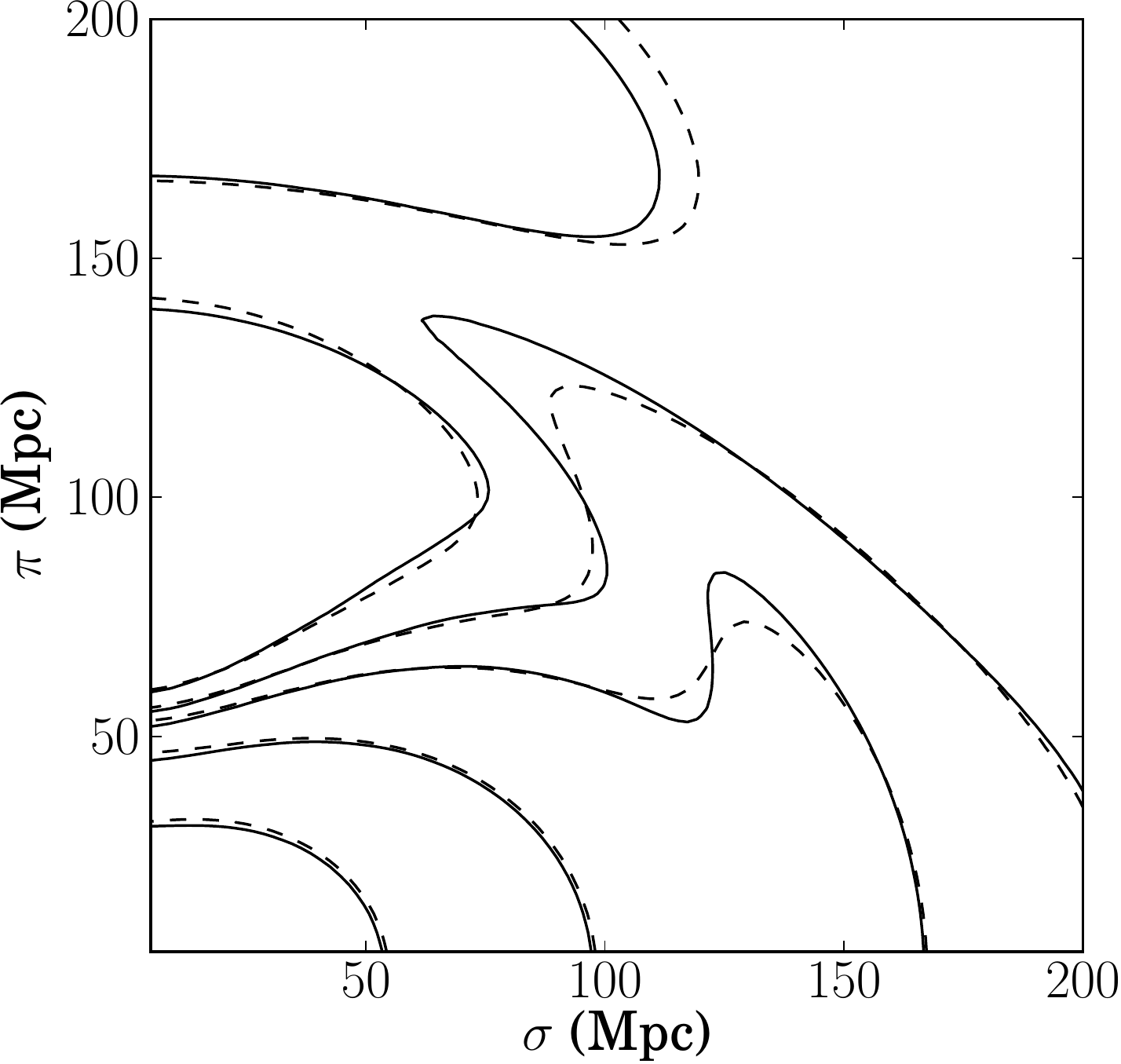}
\includegraphics[width=1\columnwidth]{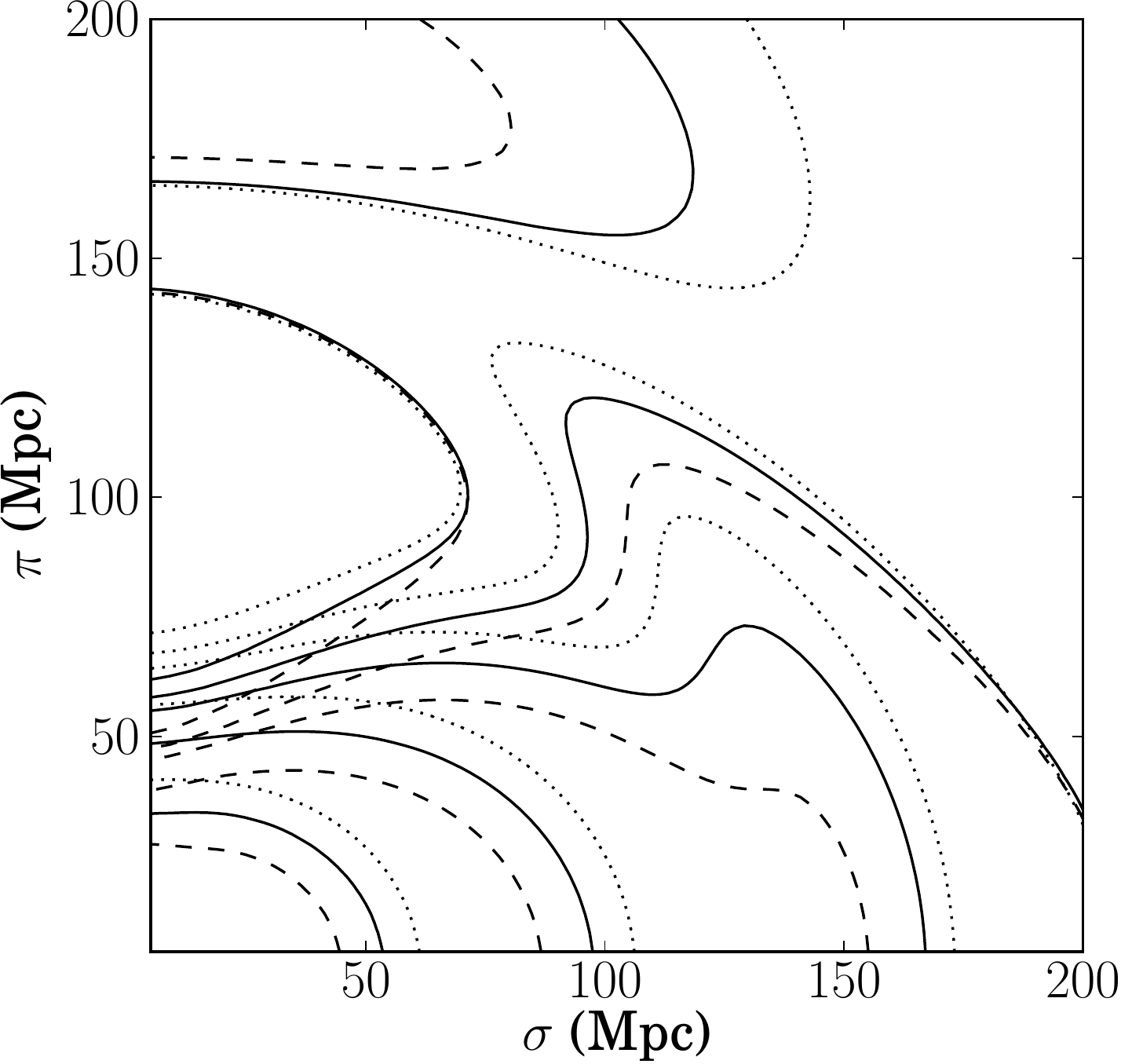}\\
\includegraphics[width=1\columnwidth]{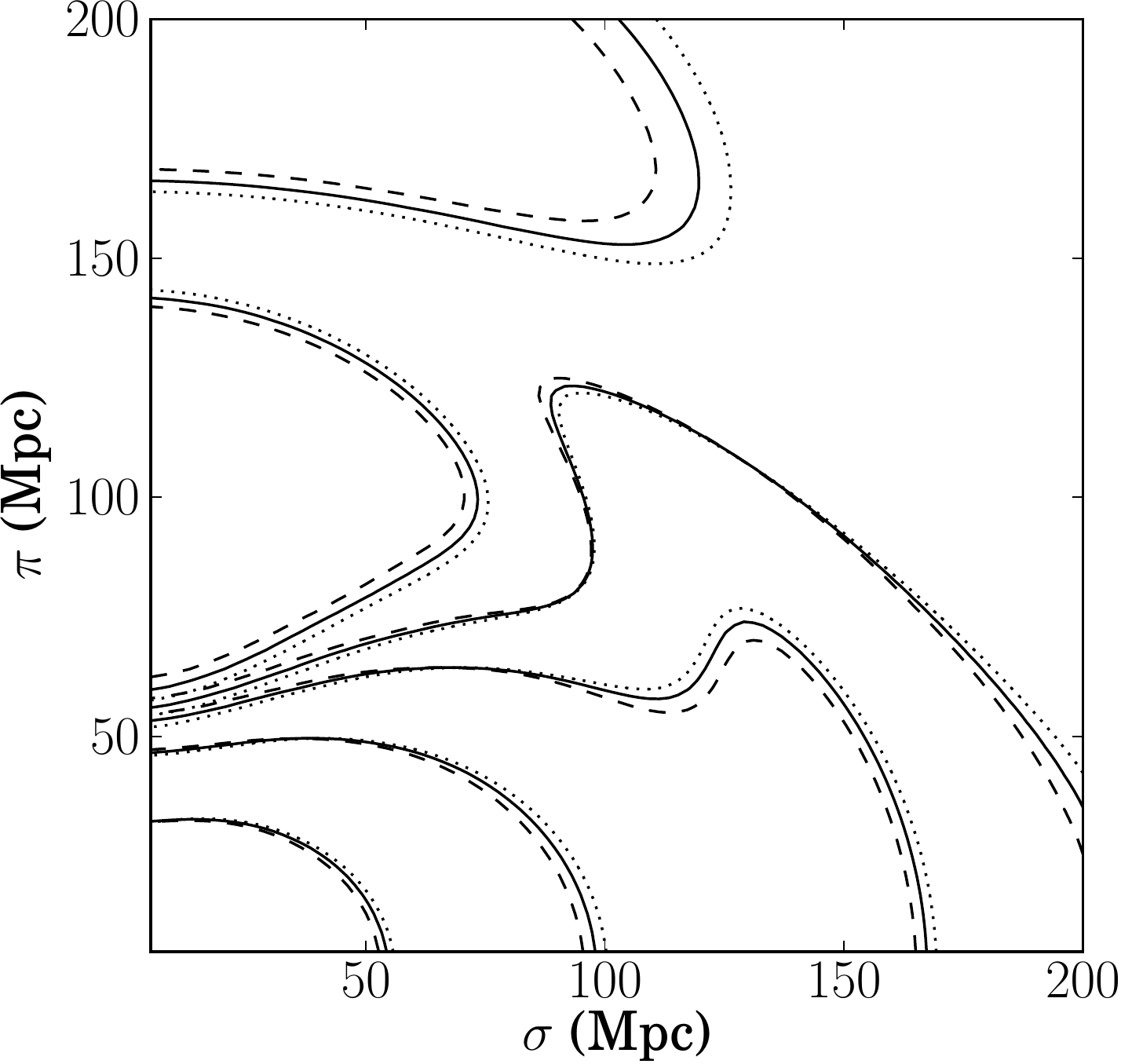}
\includegraphics[width=1\columnwidth]{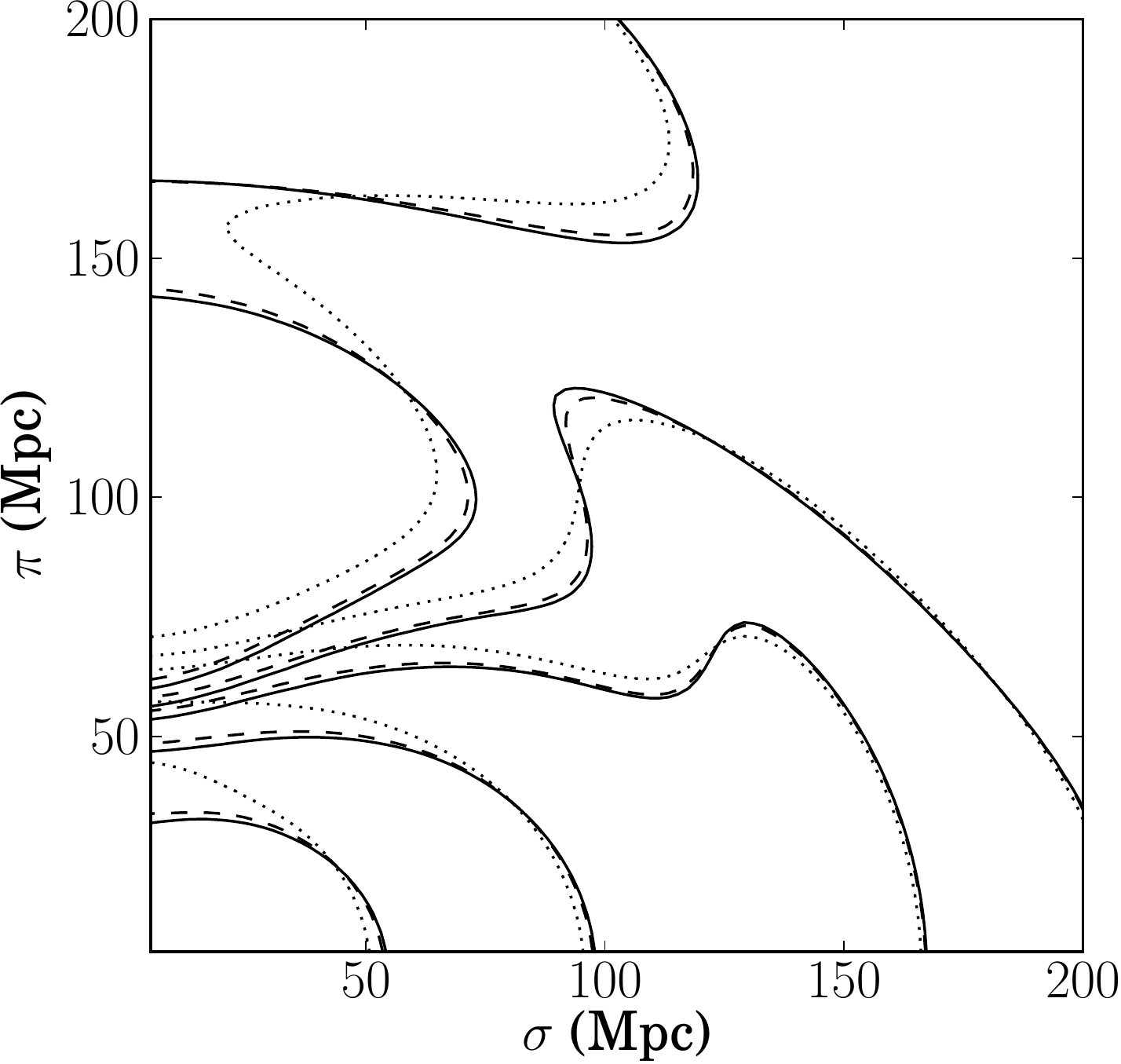}
\caption{\label{fig:xi_sys} ({\em Top left panel}) We compare $\xi_s(\sigma,\pi)$ calculated using linear theory (solid contours) and non--linear theoretical model (dash contours). The levels of contours are given by $(0.1, 0.02, 0.005, 0.001, -0.003)$. ({\em Top right panel:}) We show equi-clustering contours for 3 bias models with $b=1.5, 2.0, 2.5$ represented by dotted, solid and dash contours respectively.  ({\em Bottom left panel}) Equal clustering contours for three values of $\Omega_{\Lambda}$. The values  $\Omega_{\Lambda}$ = 0.62, 0.68, 0.73 correspond to the dotted, solid and dashed lines respectively. ({\em Bottom right panel:}) Same as the left panel but now we plot 3 different FoG models with $\sigma_v=2, 8, 11$ Mpc, represented by solid, dash and dotted contours.}
\end{figure*}

In the top right panel of Fig.~\ref{fig:xi_sys}, we present the $\xi(\sigma,\pi)$ with variation of bias. The solid, dash and dotted contours represent the $\xi(\sigma,\pi)$ with $b_0=1.5$, 2.0 and 2.5 respectively. No significant deviation of BAO ring is observed with variation of coherent galaxy bias $b$. We present it more quantitatively in the table below. The measured $D_A$ and $H^{-1}$ does not deviate more than $0.8\%$ with respect to the variation of $b_0$. The fractional deviation is presented in the bracket. Therefore we are able to measure distances through BAO ring despite the unknown galaxy bias. 

\begin{center}
\begin{tabular}{lllll}
bias & $D_A(\mpc)$ & $H^{-1}(\mpc)$ &   \\
\hline
 1.5 & 1395.18  ( 0.00 \%) & 3241.28 ( 0.20\%) &  \\
 2.0 (fid) & 1395.18  ( 0.00 \%) & 3234.76  ( 0.00 \%)&  \\
 2.5 & 1384.29 ( -0.78\%) &3234.76 ( 0.00\%)&  \\
\hline
\end{tabular}
\end{center}

\subsection{Standard ruler against variation of cosmological model}
The primeval matter-energy fluctuations are assumed to be known by CMB experiments. We test whether it is sufficient for determining distances through the observed BAO peaks. In the context of standard cosmology, the shape of spectra is determined before the last scattering surface, and in linear theory, it evolves coherently through all scales. The history of structure formation evolution is divided into two regimes; epochs before matter--radiation equality and a later epoch of coherent evolution of unknown effect on structure formation from new physics. 

True cosmological model can be different from the fiducial model, but it satisfies with the given CMB prior. If we lives in $\Lambda$CDM universe, but the fiducial cosmology is not consistent with the true cosmology. There is only one degrees of freedom to vary with the imposed CMB priors. Here we vary $H_0$ to remain $\omega_b$ and $\omega_m$ unchanged. Two different cosmologies with the same BAO structure are tested in the table below. We test $H_0=61\,{\rm km/s}/\mpc$ and $73\,{\rm km/s}/\mpc$ cases. If those cases are considered as the fiducial model, we would not find any difference in the measured distances, within $0.2\%$. The BAO indeed remains as a standard ruler against variation of cosmological model, as far as those stem from the same primeval spectra at the last scattering surface. It suggests that BAO will probe the true distances regardless of the fiducial models used.

\begin{center}
\begin{tabular}{l|lll}
$\Omega_{\Lambda}$ & $D_A$ (\% fiducial)& $H^{-1}$ (\% fiducial)&   \\
\hline
0.62 &  1493.08 ( -0.19\%) & 3396.97 ( 0.20\%) &  \\
0.68 & 1395.18  ( 0.00 \%) & 3234.76  ( 0.00\%)&  \\
0.73 & 1395.18  ( 0.00 \%) & 3234.76  ( 0.00\%)&  \\
\hline
\end{tabular}
\end{center}

In the bottom left panel of Fig.\ref{fig:xi_sys} we show the clustering contours for $\Omega_{\Lambda}=0.62$, 0.73. We see that variations in  $\Omega_{\Lambda}$ lead to shifts in the contours but as we determined and presented in the table above, the BAO ring shape is not affected.

\subsection{Non--perturbative effect from randomness of peculiar velocity: Weak limit}
The contamination due to the randomness of velocity is not formulated in the above perturbation theory. This effect dubbed as FoG (hereafter Finger of God) is phenomenologically given by Gaussian approximation as,
\ba
\tilde{P}(k,\mu) = \left[P_{\delta\delta}(k) + 2\mu^2 P_{\delta\Theta}(k) + \mu^4 P_{\Theta\Theta}(k)\right] G^{\rm FoG}(k\mu\sigma_v)\,,
\label{eq:TNS10}
\ea
where $G^{\rm FoG}$ is given by,
\ba
G^{\rm FoG}=\exp\left[-(k\mu\sigma_v)^2\right].
\ea
Here $\sigma_v$ denotes the velocity dispersion. 

The fiducial value of $\sigma_v$ is given by the linear velocity dispersion of $\sigma_v=5\mpc$. Once the power spectrum has been computed, it is straightforward to compute the correlation function. The redshift-space correlation function $\xi_s(\sigma,\mu)$ is generally expanded as
\ba\label{eq:xi_eq}
\xi_s(\sigma,\pi)=\int \frac{d^3k}{(2\pi)^3} \tilde{P}(k,\mu)e^{i{\bf k}\cdot{\bf s}}.
\ea
Here we also consider the bias due to galaxy clustering. The matter density field is replaced with the galaxy density field by coherent bias factor $b=2$. The systematic uncertainty due to galaxy bias is examined in the next subsection.

We first consider the fingers-of-god effect where the galaxy distribution is elongated in redshift space, with an axis of elongation pointed toward the observer. It is caused by a Doppler shift associated with the random peculiar velocities of galaxies bound in structures such as clusters. The deviation from the Hubble's law relationship between distance and redshift is altered, and this leads to inaccurate distance measurements. 

In a simple case we can assume that this FoG is given by a factorized form as in Eq.~(\ref{eq:TNS10}).
Because we are mainly interested in clustering about the BAO ring, which has the most contribution from quasi linear scale spectra, the leading order of $G^{\rm FoG}$ function is dominant over all other higher order contributions. Thus the detailed functional form of $G^{\rm FoG}$ is not so important as far as the $\sigma_v$ is assumed to be undetermined.
%
When $(k\mu\sigma_v)^2 \ll 1$, the leading order term of Eq.~(\ref{eq:TNS10}) is dominant over all other higher orders, and the estimated errors are immune from the exact functional form of Eq.~(\ref{eq:TNS10}).

We now proceed to check if the FoG distortion effects the BAO peak position. In the table below we show the derived distance measurements using models with various $\sigma_v$ choices, of 2, 5, 8, 11, 15 $\mpc$. The theoretical $\xi(\sigma,\pi)$ is presented in the bottom right panel of Fig.~\ref{fig:xi_sys}. It is difficult to find the deviation from the figure. However, the measured $H^{-1}$ is possibly biased by a couple of percentage level. When we consider the low resolution map such as BOSS, it would not be much influential. But when we fit the high resolution experiment such as DESI, we have to marginalise FoG effect. For the measured $D_A$, it is again immune from the FoG effect. The detailed results are presented below.

\begin{center}
\begin{tabular}{lllll}
$\sigma_v(\mpc)$ & $D_A$ (\mpc)& $H^{-1}$ (\mpc)&   \\
\hline
2   & 1392.47  ( -0.19 \%) & 3253.96  ( 0.59\%) &  \\
5 (fid)  & 1395.18  ( 0.00 \%) & 3234.76  ( 0.00 \%)&  \\
8   & 1395.18  ( 0.00 \%) & 3234.76  ( 0.00 \%)&  \\
\hline
11 & 1397.99  ( 0.20 \%) & 3166.40 ( -2.11\%)&  \\
15 & 1397.99  ( 0.20 \%)  & 3077.53 ( -4.86\%) &  \\
\hline
\end{tabular}
\end{center}

Note that the transverse distance is measured precisely, regardless of the FoG effect. In terms of $D_A$, the BAO remains as a standard ruler independent of the random velocity effect. However, the determination of radial distance becomes uncertain due to the FoG. In the weak contamination limit of $\sigma_v\la 10\mpc$, the systematic uncertainty is smaller than a couple of percent. When we consider a lower resolution map such as the BOSS catalog, the BAO remains as a standard ruler for the radial distance as well. However, in the strong contamination limit, with $\sigma_v\ga 10\mpc$, the determination of $H^{-1}$ should  not be trustable. We will discuss this further in the following section. 

\subsection{Non--perturbative effect from randomness of peculiar velocity: Strong limit}

\begin{figure}
\centering
\includegraphics[width=1.0\columnwidth]{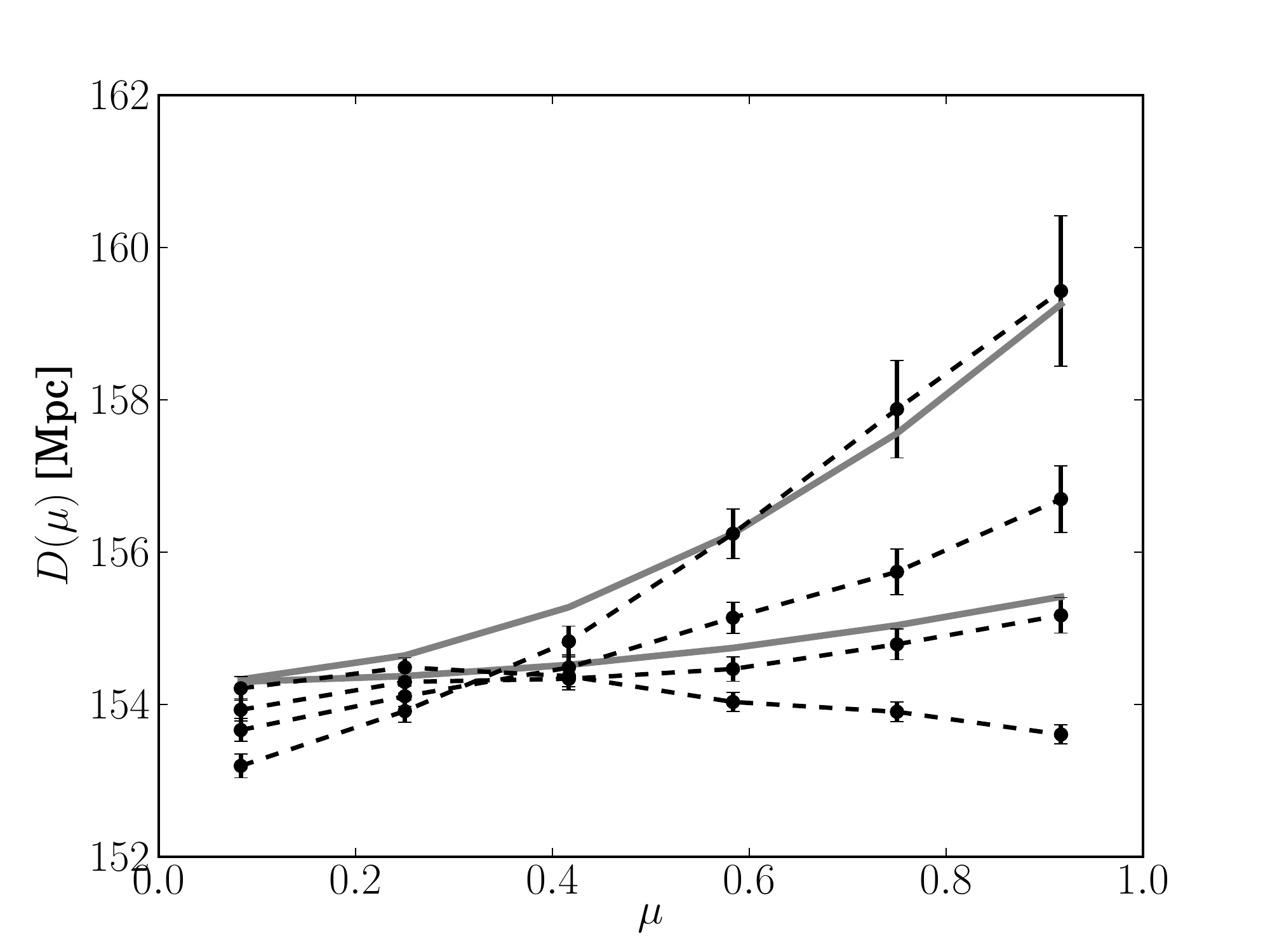}
\caption{\label{fig:D_mu} The $D(\mu)$ curve for 4 values of $\sigma_v=2, 8, 11, 15$ Mpc (4 black dashed lines). The grey sold lines are for 1 and 4\% increase in $H^{-1}$.}
\end{figure}

The measured distances using the full RSD anisotropic analysis are significantly contaminated by a systematic caused by the random peculiar velocity of galaxies. While the precise theoretical prediction for FoG effect on RSD clustering is unknown, this contamination is marginalised through at quasi linear scales in which the FoG can be parameterised by a coefficient at leading order. The coherent motion and $H^{-1}$ are poorly determined by this systematic. In this work, we measure distances using not the full RSD anisotropy analysis, but the locations of the BAO in $\sigma-\pi$ plane. The measured distances using BAO peak statistics are less contaminated by FoG effect presented in the previous subsection. This systematic is as invisible as others, if the dispersion of random velocity is small and low resolution maps are used.

However, if the observed $\sigma_v$ is exceeding the linear velocity dispersion by a factor of 2, the radial BAO location is influenced at a level of a couple of percent, while the transverse BAO location is still immune from the FoG effect. The measured $H^{-1}$ is biased by 2\% and 4\% levels at $\sigma_v=11\mpc$ and $15\mpc$ respectively. In the future, $H^{-1}$ is expected to be measured within this precision level. 

The variation of BAO peak locations is presented as dotted curves in Fig.\ref{fig:D_mu} with different $\sigma_v=(2,8,11,15)\mpc$ from bottom to top. In the small $\sigma_v$ limit about the linear velocity dispersion of $\sigma_v=(2,8)\mpc$, the observed $D(\mu)$ is nearly independent of the FoG systematic. In the strong $\sigma_v$ limit, much greater than linear velocity dispersion of $\sigma^{\rm lin}_v=6\mpc$, the observed BAO peak locations at $\mu\rightarrow 1$ starts to be biased at a level of several percent. The dash curves represent a shift of the BAO peak location with variation of $H^{-1}$ at 1\% and 4\%, which presents an indistinguishable signature from the case of $\sigma_v$ variation. This would suggest that we may only be able to make constraints of the combined quantity $\sigma_vH^{-1}$.

We investigate the detailed relation between the measured peak positions of $D(\mu)$ and the FoG parameter $\sigma_v$. The $D(\mu)$ is given by the following fitting formulation,
\begin{equation}
\label{eq:sigma_fit}
D(\mu)=D^{fid}(\mu)+\alpha(\sigma_v)+\beta(\sigma_v)\,\mu^2.
\end{equation}
The observed curve of running $D(\mu)$ in Fig.\ref{fig:D_mu} is well reproduced by this formulation. The best fits $\alpha(\sigma_v)$ and $\beta(\sigma_v)$ are given in the following table,
\begin{center}
\begin{tabular}{l|c|c|c}
Models & $\Omega_{\Lambda}=0.68$ & Fiducial & $\Omega_{\Lambda}=0.73$\\
\hline
\hline
$\alpha(\sigma_v=2\mpc)$ &0.020&0.015&0.032\\
$\beta(\sigma_v=2\mpc)$ & 0.145& 0.086&0.089\\
\hline
$\alpha(\sigma_v=5\mpc)$ &-0.096&-0.097 &-0.125\\
$\beta(\sigma_v=5\mpc)$ &0.916&0.881&0.947\\
\hline
$\alpha(\sigma_v=8\mpc)$ &-0.297&-0.330 &-0.338\\
$\beta(\sigma_v=8\mpc)$ &2.348&2.327& 2.384\\
\hline
$\alpha(\sigma_v=11\mpc)$ &-0.536&-0.574 &-0.621\\
$\beta(\sigma_v=11\mpc)$ &4.475&4.496&4.551 \\
\hline
$\alpha(\sigma_v=15\mpc)$ &-0.944&-0.949& -1.064\\
$\beta(\sigma_v=15\mpc)$ &8.709&8.471&9.048 \\
\hline
\end{tabular}
\end{center}

The shape of $D(\mu)$ running is quadrature for both variations of $H^{-1}$ and $\sigma_v$ in terms of $\mu$, which causes the degeneracy between $H^{-1}$ and $\sigma_v$. Then we will constrain just on the combination of $H^{-1}$ and $\sigma_v$. 

But there is a possible way to break this degeneracy. Although the patterns of $\Delta D(\mu)$ with the variations of $H^{-1}$ and $\sigma_v$ are quadrature in terms of $\mu$, both show a distinct behaviours at $\mu\rightarrow 0$ limit. Whilst $\Delta D(\mu)=0$ at $\mu\rightarrow 0$ with the variation of $H^{-1}$, the $\alpha(\sigma_v)$ is non--trivial to lead $\Delta D(\mu)\ne 0$ at $\mu\rightarrow 0$. Thus, the radial distance $H^{-1}$ can be separately measured, after marginalising $\sigma_v$. 

The observed $\alpha(\sigma_v)$ and $\beta(\sigma_v)$ are presented in the above table with different cosmological models. We find that both parameters are weakly dependent on cosmology models. Thus we will use this known formulation when we marginalise the contamination by FoG systematic. 

\section{Application for CMASS catalogue of BOSS DR11}
\label{sec:mocks}

The 2D anisotropy analysis has been applied for BOSS DR11 CMASS map in our previous work, to measure cosmic distances after marginalising the large scale clustering information and the velocity randomness effect. The improved RSD theoretical model allows us to remove systematics at linear regime, and provides us with conservative, but trustable result. The measured distances are $D_A=1425.2^{+29.6}_{-32.2} \mpc$ and $H^{-1}=3223.6^{+203.4}_{-175.8} \mpc$. 
The observed fractional errors of $D_A$ and $H^{-1}$ are 2\% and 6\% respectively, which is poorer than prediction due to the conservative selection of data. The current RSD theoretical model is not yet confidently improved at small scales, and the conservative cutoff strategy removes information at $s<50\mpcoh$ and $\sigma<40\mpcoh$.

The methodology explained in this manuscript is less dependent on RSD systematics, and no information is cut off at any scale. We use 600 simulations produced for mocking the BOSS DR11 CMASS (North sample) galaxies, and compare the measured distances using BAO peaks with $D_A$ and $H^{-1}$ measurements using full 2D anisotropy analysis. The cosmology models for the simulations is $\Lambda$CDM with $\omega_b=0.022$, $\omega_c=0.12$ and $h=0.67$, which provides $D_A=1395.18\mpc$ and $H^{-1}=3234.76\mpc$ of Planck $\Lambda$CDM concordance model. The uncertainty due to galaxy random motions is parametrised by one dimensional velocity dispersion parameter $\sigma_p$, when the map is analysed using the full 2D anisotropy method. The $\sigma_p$ is significantly degenerate with linear velocity growth function in linear regime, and very poorly determined. The measured $\sigma_p$ is $9.3^{+5.4}_{-5.7}\hompc$ for BOSS DR11 CMASS. It is correspondent to the weak FoG contamination limit, but there is possible impact on the cosmic distance determination using out method, as it is observed at the edge of weak FoG bound. In this section, we exploit both weak and strong FoG contamination treatments.

\begin{figure}
\includegraphics[width=\columnwidth]{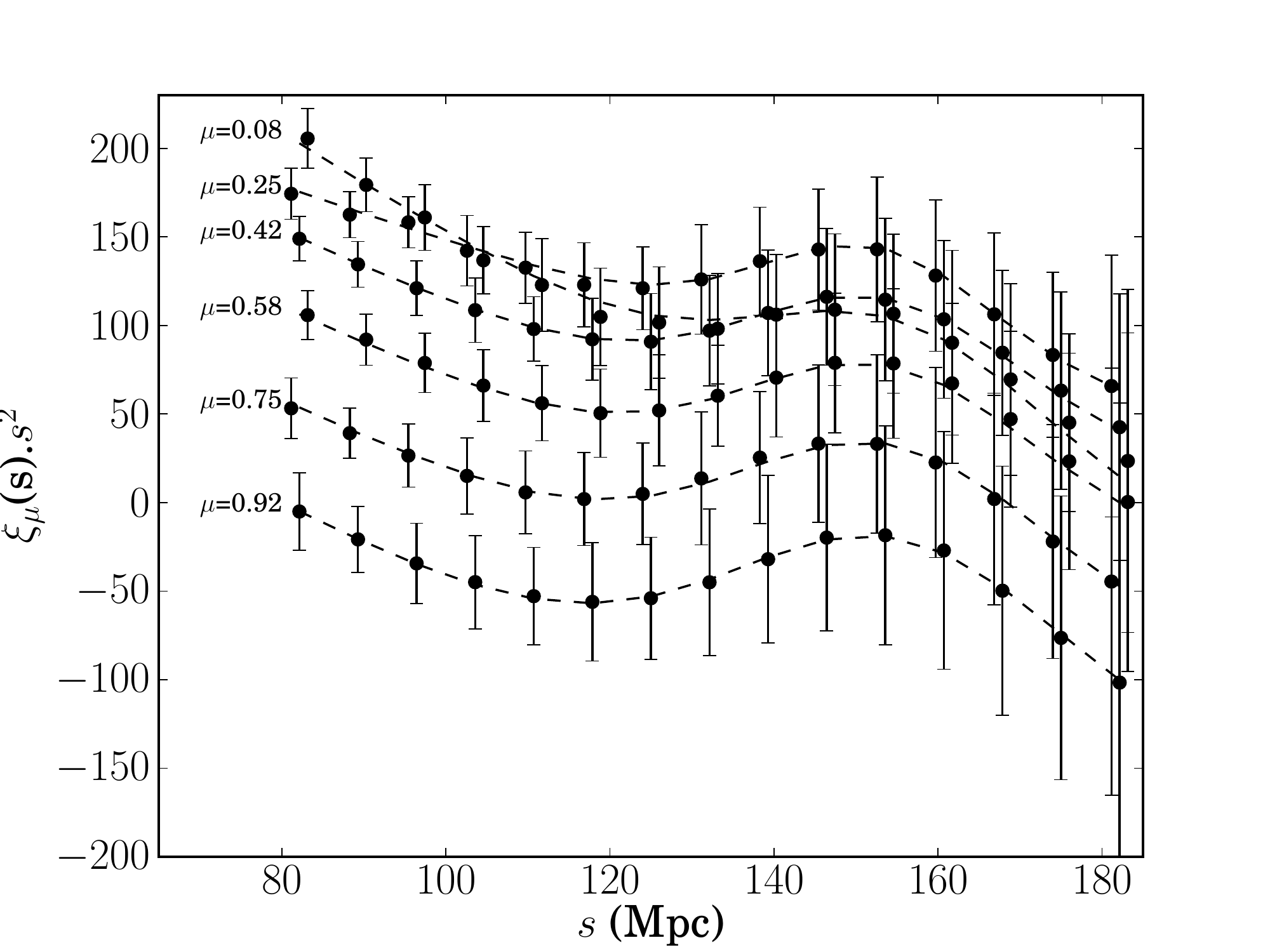}
\caption{\label{fig:xi_sim} The observed one-dimensional correlation function $\bar\xi_\mu(s)$ for CMASS simulations is presented for each wedge as black solid points. The corresponding $\mu$ for each curve is given by $\mu$=0.92, 0.75, 0.58, 0.42, 0.25, 0.08 from the bottom to top. The  dashed curves represent the best fit curves defined by Eq.~\ref{eq:model}.}s
\end{figure}

\begin{figure*}
\includegraphics[width=\columnwidth]{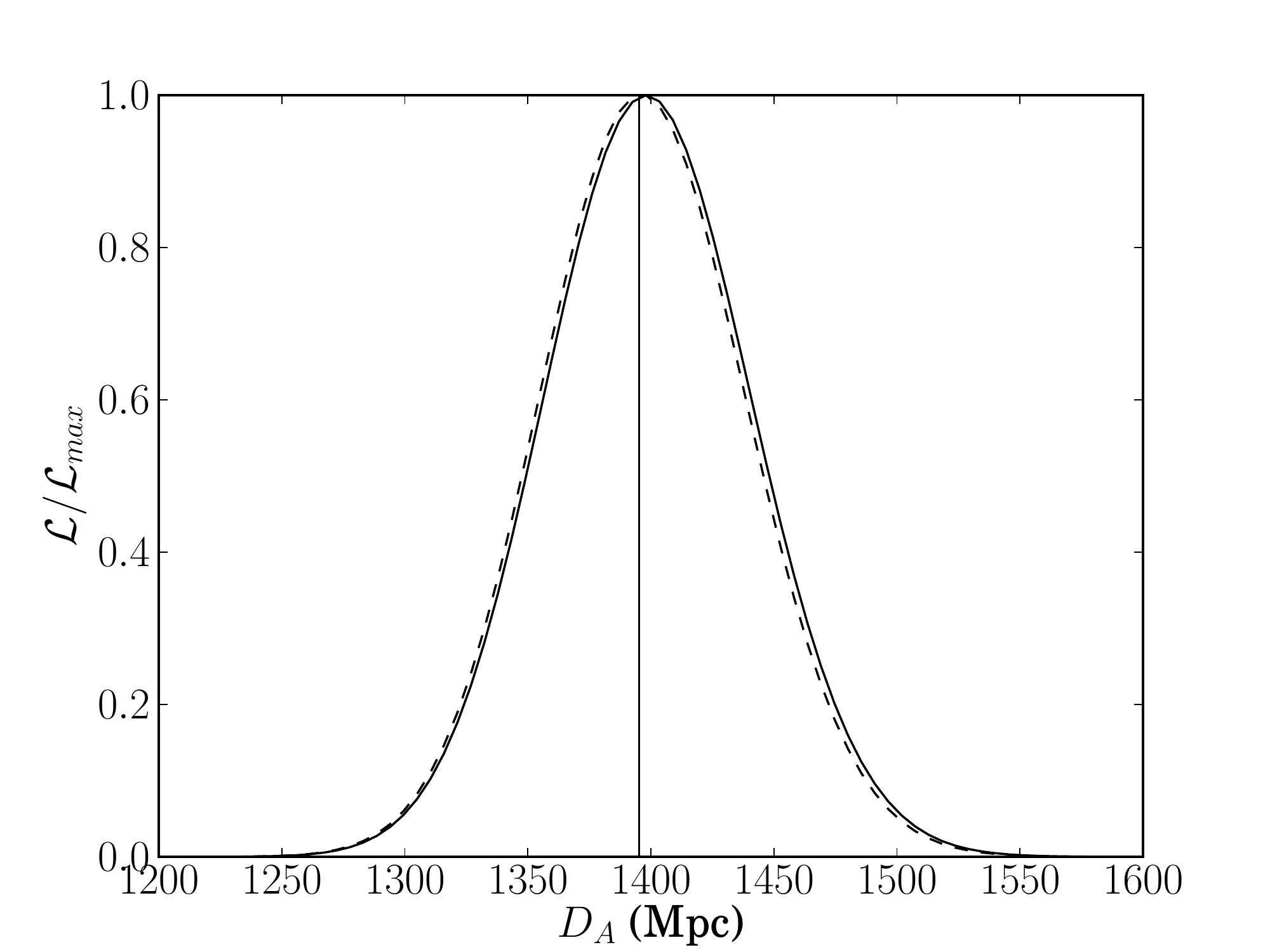}
\includegraphics[width=\columnwidth]{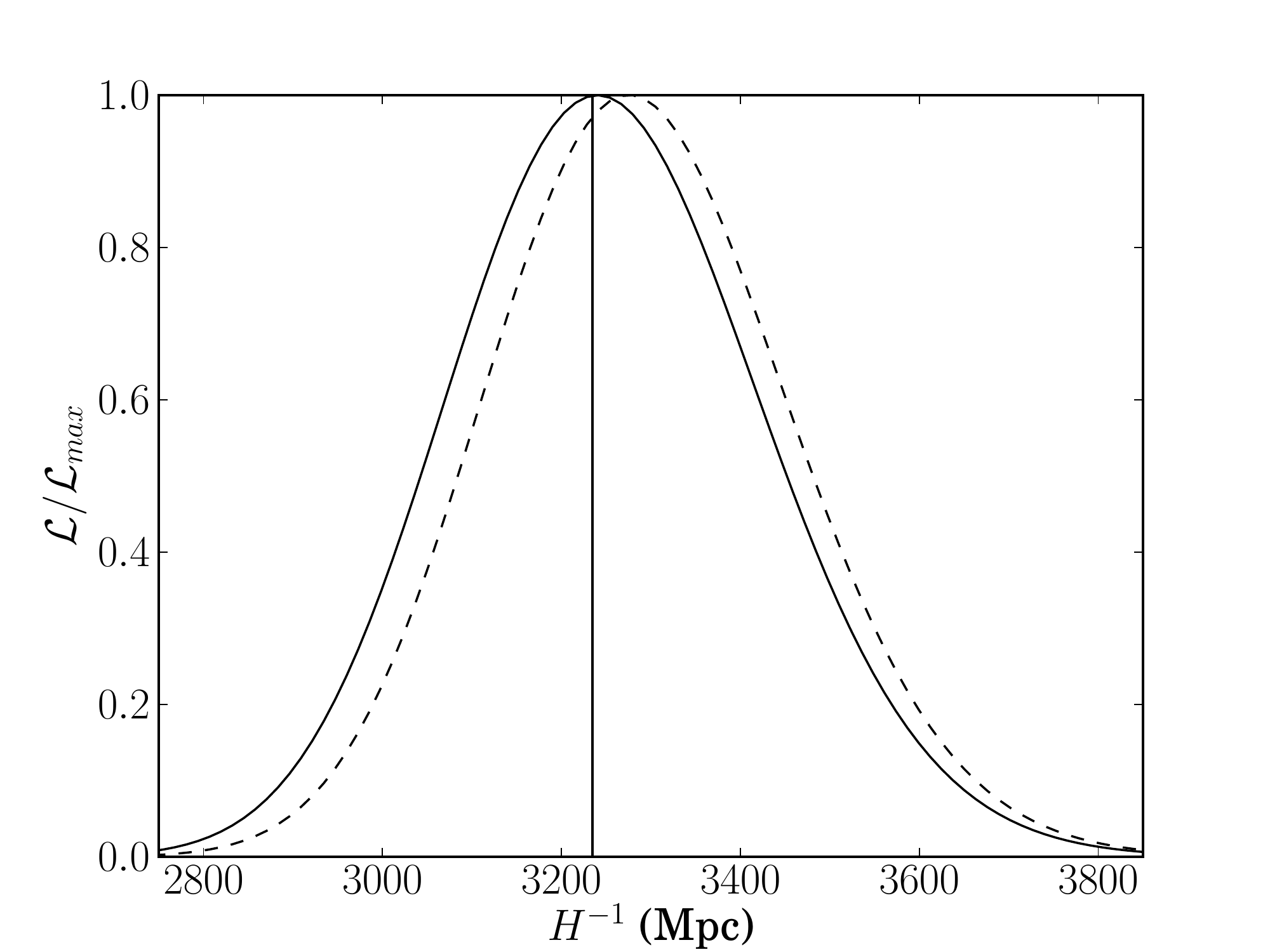}
\caption{\label{fig:Dpara_lik}  {\em Left panel:} The one dimensional relative likelihood for $D_A$, while fixing $\sigma_v=0$, is presented as the dashed curve, and the one dimensional relative likelihood, when marginalising over the unknown $\sigma_v$, is presented as the solid curve. The dark and light shaded region represent 68\% and 95\% confidence levels for the solid curve. The solid vertical line denotes the fiducial $D_A$ value known from the simulation. {\em Right panel:} the same as the in the left panel, but now focusing on the constraint of the radial distance, $H^{-1}$.}
\end{figure*}

The simulations that we use are {\tt PTHALO} mock galaxy catalogs\footnote{These mock catalogs are available from: http://marcmanera.net/mocks/index.html} created by~\citet{Manera:2012sc}, which are designed to investigate the various systematics in the galaxy sample from Data Release 11 (DR11) of the Baryon Oscillation Spectroscopic Survey (BOSS) \citep{2009astro2010S.314S,2011AJ....142...72E,2012MNRAS.427.3435A}, referred to as the ``CMASS" galaxy sample. In constructing the mock galaxy catalogs,~\citep{Manera:2012sc} utilized second-order Lagrangian perturbation theory (2LPT) for the galaxy clustering driven by gravity, which enables the creation of a mock catalog much faster than running an $N$-body simulation. The mocks catalogs constitute 600 density field realizations  which span the redshift range of the observed galaxies in our sample i.e.  $0.43<z<0.7$. Each catalog contains $\sim 7 \times 10^5$ galaxies, 90\% of which are central galaxies residing in dark matter halos of $\sim 10^{13}h^{-1}M_\odot$.

\subsection{Cosmic distances at weak FoG limit}

The 2D correlation function $\xi(\sigma,\pi)$ is calculated for the simulations as described in Eq.~\ref{eq:LSesti}, and the averages of $\xi(\sigma,\pi)$ at the given $\sigma$ and $\pi$ bins are derived using 600 realisations. The means of $\xi(\sigma,\pi)$ are transformed into correlation function $\xi(s,\mu)$. The radial coordinate bin is given at $80\mpc <s<180\mpc$ with spacing $\Delta s=7\mpc$ which encloses the approximate BAO characteristic scale of $150\mpc$. The directional coordinate $\mu$ is divided into 6 bins as defined in the previous section.

The covariance matrix of $\xi$ among different $(k,\mu)$ bins is estimated using 600 mock catalogues. This number of realisations exceeds the number of $(k,\mu)$ bins of 84, which provides us with a stable inverse matrix. Using the correlation function computed from each of the 600 mock catalogs, we estimate the covariance matrix as 
\begin{equation}
{\rm Cov}(\xi_i,\xi_j)=\frac{1}{N-1}\sum^{N}_{k=1}[\xi_k({\bf r}_i)-\overline{\xi}({\bf r}_i)][\xi_k({\bf r}_j)-\overline{\xi}({\bf r}_j)],
\end{equation}
where $N=600$, $\xi_k({\bf r}_i)$ represents the value of the correlation function of $i$th bin of $(\sigma,\pi)$ in $k$th realization, and $\overline{\xi}({\bf r}_i)$ is the mean value of $\xi_k({\bf r}\_i)$ over realizations. We can then obtain the correlation matrix as 
\begin{equation}
C_{ij}=\frac{{\rm Cov}(\xi_i,\xi_j)}{\sqrt{{\rm Cov}(\xi_i,\xi_i){\rm Cov}(\xi_j,\xi_j)}}.
\end{equation}
The inverse of $C_{ij}$ is well defined and thus does not require any de-noising procedures e.g.  singular value decomposition. The parameters in Eq.~\ref{eq:model} in all 6 $\mu$ bins are fitted simultaneously using this full covariance matrix. 

However, the covariance matrix estimated from a finite number monte carlo realisations has been shown  
to be biased (\citet{2007A&A...464..399H}, see also 
\citet{2013MNRAS.435...64K,2014MNRAS.439.2531P}). 
An unbiased covariance matrix $C$, can be obtained by multiply the original covariance $\hat{C}$ by a correction factor,
\begin{equation}
C^{-1}=\frac{N_{mocks} - N_{bins}-2}{N_{mocks}-1}\ \hat{C}^{-1}\ ,
\end{equation}
as derived by \citet{2007A&A...464..399H},
where $N_{bins}$ is the number of measurement bins of any statistic, in our case $\xi(\sigma,\pi)$, used for the 
analysis. 

The wedge correlation function is presented in Fig.~\ref{fig:xi_sim}. The corresponding $\mu$ for each wedge curve is given by $\mu$=0.92, 0.75, 0.58, 0.42, 0.25, 0.08 from the bottom to the top. The measured average points of $\xi(s,\mu)$ from 600 mocks are represented as circles points, and the errors are computed using the full covariance of $C_{ij}$. The BAO peak locations for all wedge correlation functions appear consistent with each other, and they are found using the same fitting formulation in Eq.~\ref{eq:model}.

We analyse the measured wedge points in Fig.~\ref{fig:xi_sim}, and determine the transverse and radial BAO characteristic scales as $D_{\perp}=155.5^{+7.9}_{-7.9}\mpc$ and $D_{||}=154.8^{+4.2}_{-4.8}\mpc$. These measurements are transformed into cosmic distances $D_A= 1400.2^{+38.4}_{-43.8}\mpc$ and $H^{-1}= 3263.4^{+165.4}_{-165.4}\mpc$. The transverse and radial cosmic distances are estimated with about 3\%  and 5\% fractional errors, respectively. Both errors are nearly equivalent to the observed fractional errors for real BOSS CMASS catalogue, of which cosmic distances are computed using the full anisotropy analysis. This methodology provides remarkable precision for cosmic distance determination using BAO peaks only, considering that only the measured correlation points around the BAO peak location are exploited.

\subsection{Correction using FoG contamination treatment}

The dashed curves in the left and right panels of Fig.~\ref{fig:Dpara_lik} represent the the relative likelihood functions. The fiducial cosmic distances are shown as solid vertical lines in both panels, which are well enclosed within those likelihood function. We compare the measured transverse distance $D_A=1400.2\mpc$ with the fiducial one $1395.2\mpc$. The measurement is biased by $\Delta D_A=5\mpc$ which is correspondent to 12\% deviation against 1--$\sigma$ deviation for $D_A$, which is not considered to be significant. But the observed bias for the radial distance is $\Delta H^{-1} = 28.6 \mpc$ which is about 17\% off-set against the 1--$\sigma$ deviation for $H^{-1}$.

We are interested in understanding any possible systematic for measuring $H^{-1}$ using our method, despite the small deviation to be ignored. As the more significant bias is observed in the radial direction, it can be caused by FoG contamination. The galaxies in BOSS CMASS catalogues are not expected in the strong FoG regime, but the measured $\sigma_v$ from the full anisotropy analysis is placed in the boundary between weak and strong FoG regimes. Therefore, it would be interesting to apply our FoG contamination treatment, and present the measured cosmic distance after marginalising FoG systematic. The measurements are presented as solid curves in Fig.~\ref{fig:Dpara_lik}. The dark and light shaded regions represent 65\% and 95\% confidence levels respectively. The measured transverse distance is $D_A= 1397.0^{44.3}_{33.2} \mpc$ of which deviation from the fiducial value is $\Delta D_A = 1.8\mpc$. It is biased only 4\% against the 1--$\sigma$ deviation of $D_A$. The measured radial distance is $H^{-1}= 3217.5^{+179.9}_{-154.2}\mpc$, and $\Delta H^{-1}=17.3\mpc$ which is only 10\% deviation from the 1--$\sigma$ deviation of $H^{-1}$. This correction is represent as the solid curves in Fig.~\ref{fig:Dpara_lik}. After accounting for he FoG systematic, the results are improved significantly.

\begin{figure}
\centering
\includegraphics[width=0.93\columnwidth,   bb=59 21 473 411]{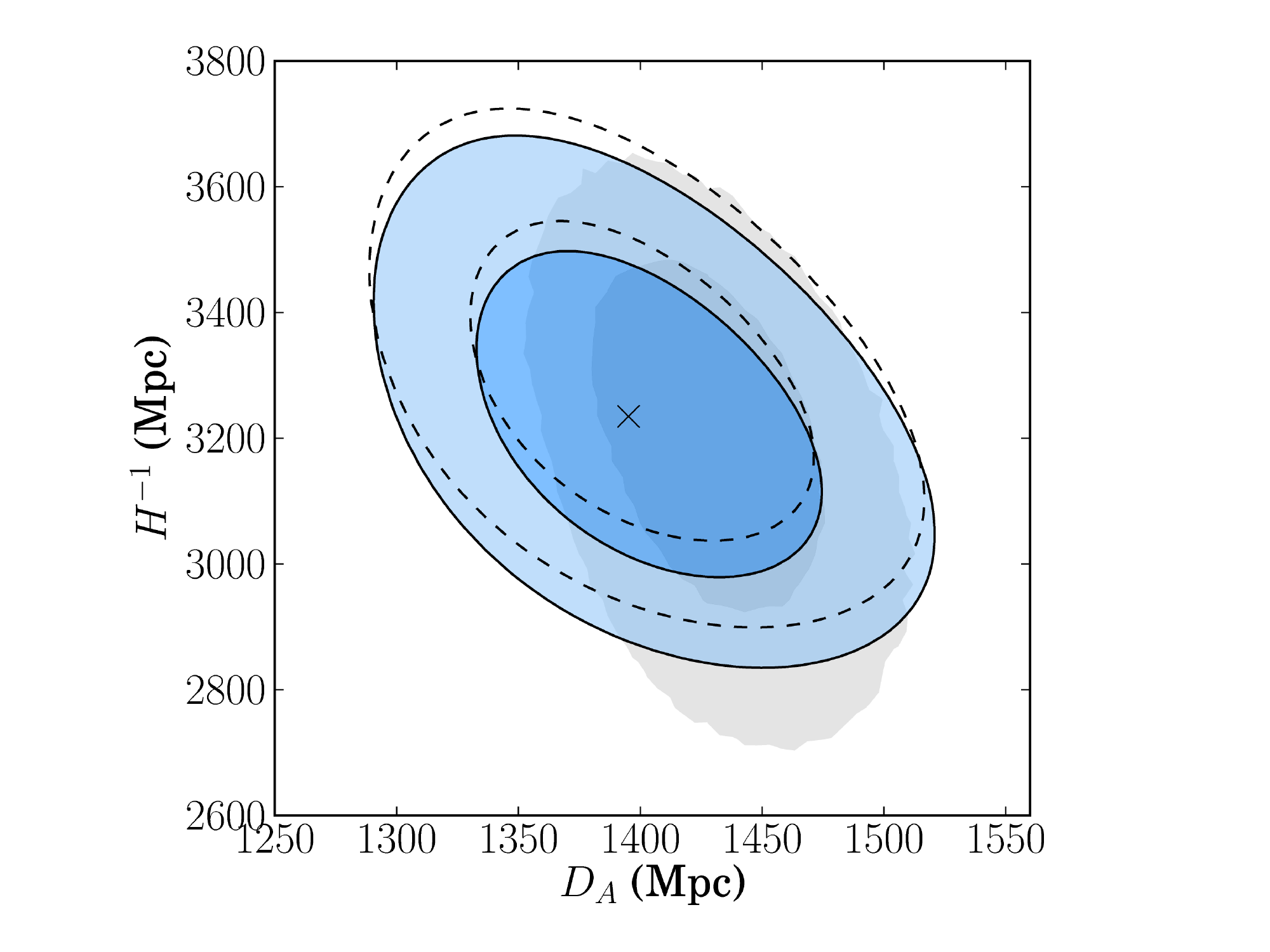}
\caption{\label{fig:con_mock}  The 2-dimensional contours in the $D_A$-$H^{-1}$ plane for the case of fixed  $\sigma_v$ (solid line) and after marginalising over  $\sigma_v$ (dashed line). The grey shaded regions are results obtained from \citet{2014arXiv1407.2257S} using the full anisotropic correlation function of the BOSS DR11 CMASS catalogue. Contours are quoted at the 1 and 2-$\sigma$ level. }
\end{figure}

In Fig.~\ref{fig:con_mock} we show the 1 and 2-$\sigma$ error bounds on $D_A$ and $H^{-1}$ for the marginalised $\sigma_v$ case (solid lines and blue shaded region) and the fixed FoG case, $\sigma_v=0$ (dashed contours). The grey shaded regions are results obtained from \citet{2014arXiv1407.2257S} using the full anisotropic correlation function of the BOSS DR11 CMASS catalogue, which we include for a comparison of the size of the constraints only as the mean values need not be consistent. We find that the containing power of our methodology is are consistent with other works based on the full shape of the anisotropic clustering. 
For instance in \citet{2014MNRAS.441...24A} using the power spectrum and correlation function with the reconstruction method recover $D_A$ to 1.4\% precision compared to our result of 2.8\% and they find $H^{-1}$ with $3.4\%$ precision to compared to our $5.3\%$. Although our constraints are slightly larger the methodology we use is quite strictly model independent and fairly straight forward to implement.

\section{Conclusions}

In our analysis we have focused on the large scale clustering of galaxies in an attempt to immunise ourselves from the uncertainty in small-scale non-linear physics. Thus we have chosen to work in the range 80-180 Mpc. Since working in these large scales, our clustering statistics carry little information of the growth of structure and this was indeed our aim as we concentrate on maximising the information gained from the Alcock-Paczynski geometrical distortions. 

We have investigated the sensitivity of the shape of the BAO ring to various systematics. We find that the shape of the BAO ring is invariant to galaxy bias and unknown shape change in the primordial power spectrum. However  non-linearities in the density field and non-linear FoG distortions contaminate the estimated BAO scales and thus the recovered values of $D_A$ and $H^{-1}$. Fortunately, the contamination of the FoG effect is not perfectly degenerate with the geometrical distortions and thus in future sub-percent level precision experiments the two effects can be disentangled. Thus we hope in that future measurements will allow us to focus on measurements of the AP effect and to infer cosmological parameters pertaining to the expansion history.

Using mock data comparable with the BOSS DR11 CMASS catalogue we show that we can obtain constraints on $D_A$ and $H^{-1}$ at the level of $\sim 3\%$ and $\sim 5\%$ respectively. We tested our methodology using mock galaxy catalogs and found that we can recover the input cosmology to a level of precision, competitive with other results in the literature. In future work we plan to further test and apply the methodology we have developed here to the observational data of SDSS BOSS \citep{OhSabiuSong}.

\section*{Acknowledgements}
Data analysis was performed using the high performance computing cluster at the Korea Astronomy and Space Science Institute.

{}

\end{document}